\documentclass[aps,pra,amsmath,amssymb]{elsart}
\usepackage{graphicx}% Include figure files
\usepackage{dcolumn}% Align table columns on decimal point
\usepackage{bm}% bold math

\begin{document}

\begin{frontmatter}
\title{Coherent matter wave inertial sensors for precision measurements in space}
\author[IOTA]{Y. Le Coq,}
\author[IOTA]{J.A. Retter}
\author[THALES]{S. Richard}
\author[IOTA]{A. Aspect}
\author[IOTA]{P. Bouyer }
\address[IOTA]{Groupe d'Optique Atomique, Laboratoire Charles-Fabry
de l'Institut d'Optique,\\ UMRA 8501 du CNRS, B\^{a}t.\ 503,
%Campus Universitaire d'Orsay,
B.P.\ 147, 91403 ORSAY, France}
\address[THALES]{Thales TRT France, 
Domaine de Corbeville, 91404 ORSAY, France}

\date{\today}

\begin{abstract}
We analyze the advantages of using ultra-cold coherent sources of
atoms for matter-wave interferometry in space. We present a
proof-of-principle experiment that is based on an analysis of the
results previously published in \cite{Richard2003} from which we
extract the ratio $h/m$ for $^{87}$Rb. This measurement shows that
a limitation in accuracy arises due to atomic interactions within
the Bose-Einstein condensate.
Finally we discuss the promising role of coherent-matter-wave
sensors, in particular inertial sensors, in future fundamental
physics missions in space.
\end{abstract}

%\pacs{PACS numbers:
  %    {03.75.-b}, %{Matter waves}
     % {32.80.Pj} %{Optical cooling of atoms; trapping}
     %} % end of PACS codes
%

\end{frontmatter}

%%%%%%%%%%%%%%%%%%%%%%%%%%%%%%%%% body %%%%%%%%%%%%%%%%%%%%%%%%%%%%%%%%%
%

% [1 intro]
%

Atom interferometry \cite{Clauser:1988,Borde91,Chu91,Pritchard91,Berman}
has long been one of the most promising candidates for
ultra-precise and ultra-accurate measurement of gravito-inertial
forces
\cite{Peters97,Gustavson97,Gustavson00,Peters99,Snadden98,Borde02,Fattori:2003}
or for precision measurements of fundamental constants
\cite{wicht2001}. The realization of Bose-Einstein condensation
(BEC) of a dilute gas of trapped atoms in a single quantum state
\cite{Anderson95,Davis95,Bradley95} has produced the matter-wave
analog of a laser in optics \cite{Mewes97,Anderson98,Hagley99,Bloch99}. As lasers have revolutionized optical
interferometry \cite{Cho85,Ste95,Stedman97},  so it is expected
that the use of Bose-Einstein condensed atoms will bring the
science of atom optics, and in particular atom interferometry, to
an unprecedented level of accuracy \cite{Bouyer:1997,Gupta:2002}.
In addition, BEC-based coherent atom interferometry would reach its full
potential in space-based applications where micro-gravity will
allow the atomic interferometers to reach their best performance
\cite{Hyper}.

In this document, we discuss the prospects of using atom-lasers in
future space missions to study fundamental physics. We point out
that atomic ensembles at sub-microKelvin temperatures will be
required, if the sensitivity of space-based atom-interferometers
is to reach its full potential.  In addition, we show that
interactions within a Bose-Einstein condensate (or atom laser)
will limit the measurement accuracy of such devices and must be
controlled to a very high level. We demonstrate the latter in a
ground-based measurement, by re-analyzing the Bragg-spectroscopy
data of \cite{Richard2003} to extract the ratio $h/m$ of the
Planck constant $h$ to the atomic mass $m$ of $^{87}$Rb. This
ratio $h/m$ is related to the fine structure constant $\alpha$
\cite{alpha}, of which precise knowledge is essential for testing
the validity of measurements related to different branches of
physics (QED, solid state physics, . . .)
\cite{alpha1,alpha2,alpha3,alpha4,alpha5,alpha6}. A measurement of
$h/m$ has been proposed as a candidate microgravity mission in the
HYPER \cite{Hyper} and ICE \cite{Ice} programs, as a follow-up to
the state-of-the art measurements on earth using cold atoms
\cite{Chu,Biraben}.

\begin{figure}[h!]
\begin{center}
\includegraphics[width=3.5in]{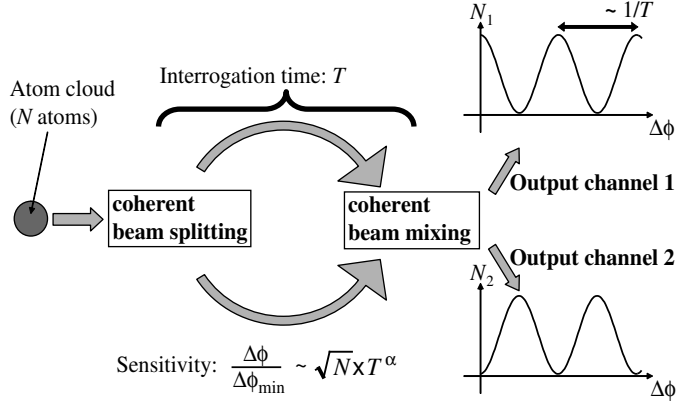}
\end{center}
\caption{Principle of an atom-interferometer. An initial atomic
wavepacket is split into two parts by the first beam splitter. The
wavepackets then propagate freely along the two different paths
for an ``interrogation time'' $T$, during which the two
wavepackets can accumulate different phases. A second pulse is
then applied to the wavepackets so that the number of atoms at
each output is modulated with respect to this phase difference.
The maximum sensitivity achievable for such an apparatus can be
defined by comparing  the variation of the number of atoms $\Delta
N$ due to the phase difference $\Delta \phi$ at the output
($\Delta N \sim N\Delta \phi /2\pi \propto N T^{\alpha}$) with the quantum
projection noise arising from atom counting $\sqrt{N}$. It scales
as $\sqrt{N}\times T^{\alpha}$.}\label{fig:Interf_principle}
\end{figure}

% [2 general principle of atom interferometry]
%
Generally, atom interferometry is performed by applying two
successive \emph{coherent} beam-splitting processes separated by a time $T$ to an ensemble
of particles (see Figure \ref{fig:Interf_principle})
\cite{Borde:1989,Giltner:1995}, followed by detection of the
particles in each of the two output channels. The interpretation
in terms of matter waves follows from the analogy with optical
interferometry. The incoming matter wave is separated into two
different paths by the first beam-splitter. The accumulation of
phases along the two paths leads to interference at the second
beam-splitter, producing complementary probability amplitudes in
the two output channels
\cite{CCT/College:1992,Antoine:2003,Storey:1994}. The detection
probability in each channel is then a sine function of the
accumulated phase difference, $\Delta \phi$.

% [3 previous work in atom interferometry]
%
Atomic clocks \cite{Kasevich:1989,Clairon:1991,Sortais:2001} can
be considered the most advanced application of atom
interferometry. In this ``interferometer'', the two different
paths of Figure \ref{fig:Interf_principle} consist of the free
evolution of atoms in different internal states with an energy
separation $\hbar\Delta\omega$. An absolute standard of time is
then obtained by servo-locking a local oscillator to the output
signal of the interferometer, which varies as $\cos
(\Delta\omega\times T)$. Atom interferometers can also be used as
a probe for gravito-inertial fields. In such applications, the
beam-splitters normally consist of pulsed near-resonance light
fields which interact with the atoms to create a coherent
superposition of two different \emph{external} degrees of freedom,
by coherent transfer of momentum from the light field to the atoms
\cite{Clauser:1988,Borde:1989}. Consequently, the two
interferometer paths are separated in space, and a change in the
gravito-inertial field in either path will result in a
modification of the accumulated phase difference. Effects of
acceleration and rotation can thus be measured with very high
accuracy. To date, ground-based experiments using atomic
gravimeters (measuring acceleration) \cite{Peters97,Peters:2001},
gravity gradiometers  (measuring acceleration gradients)
\cite{Snadden98,Mcguirk:2002} and gyroscopes
\cite{Gustavson97,Gustavson00} have been realized and proved to be
competitive with existing optical \cite{laser gyros} or
artifact-based devices \cite{cornercubes}.

%

% [4 sensitivity]
%

The ultimate phase-sensitivity of an atom interferometer is, aside
from technical difficulties, limited by the finite number of
detected particles $N$ and scales as $\Delta \phi_{\rm{min}} =
2\pi / \sqrt{N}$ (quantum projection noise limit \cite{Salomon,Wineland}). Of course, the
relation between the relative phases accumulated along the two
different paths and the actual physical property to be measured is
a function of to the ``interrogation'' time $T$ spent by the
particles between the two beam-splitters. Thus, the ideal
sensitivity of an atom interferometer is expected to scale\footnote{An
atomic clock or an atomic gyrometer, for example, has a
sensitivity proportional to $T$ and on-ground gravimeter has a
sensitivity proportional to $T^2$ due to the quadratic nature of
free-fall trajectory in a constant gravitational field.} as
$\sqrt{N}\times T^{\alpha}$ with $\alpha > 0$, and it
is obviously of strong interest to increase these two factors.

%

% [5 accuracy : necessity of free-fall]
%
Nevertheless, in practice, the \emph{absolute accuracy} of an
atom-interferometer is limited by uncontrolled, environmental
phase shifts in the interferometer, for example, due to stray
electromagnetic fields or mechanical vibrations. These residual
phase shifts must therefore be controlled and measured to better
than the desired accuracy. This is usually best achieved by
keeping these shifts as small as possible, using passive isolation
and active feedback \cite{Keith:1991}. Such constraints forbid in general the
use of external fields as a means of controlling the position of
the atoms during the phase accumulation period. In addition, any
\emph{inhomogeneity} in an external potential applied to the atoms
would usually result in a loss of coherence, decreasing the sensitivity
and dynamics of the atom interferometer. As a consequence,
most high-precision atom interferometers require that the atoms are in
free-fall between the two beam-splitting processes.
%

% [6 advantage of space]
%
Seeking to increase the sensitivity of on-ground atom
interferometers by increasing the interrogation time $T$, one soon
reaches a limit imposed by gravity. With the stringent
requirements of ultra-high vacuum and a very well controlled
environment, current state-of-the-art experimental apparatus does
not allow more than a few meters of free-fall, with corresponding
interrogation times of the order of $T \sim 200\,$ms. Space-based
applications will enable much longer interrogation times to be
used, thereby increasing dramatically the sensitivity and accuracy
of atom interferometers \cite{Hyper}.

\begin{figure}
\begin{center}
\includegraphics[width=3.5in]{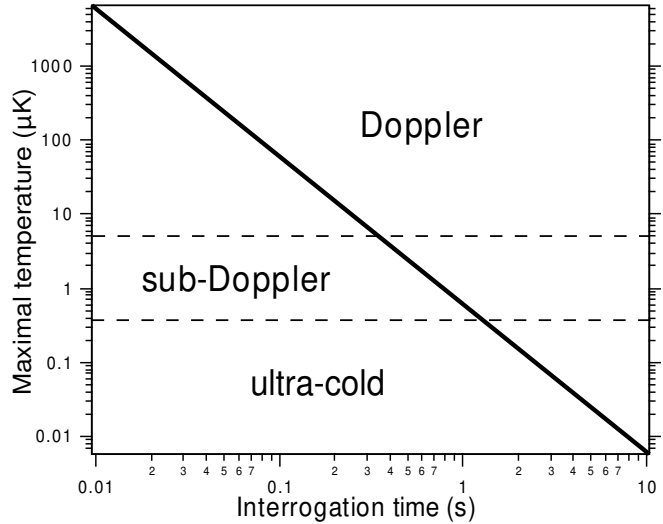}
\end{center}
\caption{Maximum temperature of atom source for a given
interrogation time. The maximum interrogation time for a given
initial temperature has been calculated for a detection area of
10\,cm$^2$ and defined as the time at which half of the atoms are
no longer detected. The dashed lines indicate the limits of
Doppler and sub-Doppler cooling. Interrogation times of several
seconds are compatible only with clouds of atoms at ultra-cold
temperatures, close to the quantum degenerate
regime.}\label{fig:T_vs_interrogation}
\end{figure}

% [7 why ultra-cold sources in space ?]
%
%
Even in space, atom interferometry with a {\em classical} atomic
source will not outperform the highest-precision ground-based atom
interferometers that use samples of cold atoms prepared using
standard techniques of Doppler and sub-Doppler laser cooling
\cite{lasercooling}. Indeed, the temperature of such sub-Doppler
laser-cooled atom cloud is typically $\sim 1\,\mu$K ($v_{\rm
rms}\sim 1\,$cm/s). In the absence of gravity, the time evolution
of cold samples of atoms will be dominated by the effect of finite
temperature: in free-space, a cloud of atoms follows a ballistic
expansion until the atoms reach the walls of the apparatus where
they are lost. Therefore the maximum interrogation time reasonably
available for space-based atom interferometers will strongly
depend on the initial temperature of the atomic source. As shown
in Figure \ref{fig:T_vs_interrogation}, the 200\,ms limit imposed
by gravity for a 30\,cm free fall is still compatible with typical
sub-Doppler temperatures, whereas an interrogation time of several
seconds is only accessible by using an ``ultra-cold'' source of
atoms (far below the limit of laser cooling) with a temperature of
the order of a few hundred nanoKelvin.

% [8 Quantum degeneracy and atom interferometry]
%
These dense, \emph{ultra-cold} samples of atoms are now routinely
produced in laboratories all around the world. Using evaporative
cooling techniques \cite{Anderson95,Davis95,Bradley95}, one can
cool a cloud of a few $10^6$ atoms to temperatures below 100\,nk
\cite{KetterleTemp}. At a sufficiently low temperature and high
density, a cloud of atoms undergoes a phase transition to quantum
degeneracy. For a cloud of bosonic (integer spin) atoms, this is
known as Bose-Einstein condensation, in which all the atoms
accumulate in the same quantum state (the atom-optical analogy of
the laser effect in optics). A BEC exhibits long range correlation
\cite{Hag99,Blo99,stenger:1999} and can therefore be described as
a coherent ``matter wave'': an ideal candidate for the future of
atom interferometry in space. The extremely low temperature
associated with a BEC results in a very slow ballistic expansion,
which in turn leads to interrogation times of the order of several
tens of seconds in a space-based atom interferometer. In addition,
the use of such a coherent source for atom optics could give rise
to novel types of atom interferometry
\cite{Bouyer:1997,Gupta:2002,Ketterle:2004,Bouyer:2004}.

% [ Interest of atom lasers as bright sources ]
%  to be added if plenty of space (calculus of luminance)

%

\begin{figure}
\begin{center}
\includegraphics[width=2.5in]{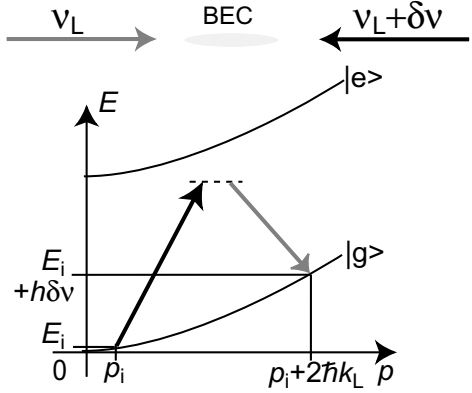}
\end{center}
\caption{Principle of Bragg scattering: a moving standing wave,
formed from two counter-propagating laser beams with a small
relative detuning $\delta \nu$, can coherently transfer a fraction
of the atoms to a state of higher momentum when the resonance
condition is fulfilled. A 2-photon Bragg scattering event imparts
a momentum $2\hbar k_{\rm{L}}$, and an energy of $h\delta\nu$ to
the atoms: thus, the first-order (2-photon) Bragg resonance for
atoms with zero initial velocity occurs at a detuning of $h \delta
\nu= 4 \hbar^2 k_{\rm{L}}^2/2m$.  This resonance condition depends
on the initial velocity of the atoms relative to the optical
standing wave.}\label{fig:Bragg2photon}
\end{figure}
%

% [ principle of Bragg scattering]
%
In our laboratory we have realized a coherent matter-wave
interferometer based on Bragg scattering \cite{Bouyer:2004}. The
principle of Bragg scattering is the following
\cite{martin:1988,kozuma:1999}: two counter-propagating laser
beams of wavevector $\pm{\bf k}_{\rm{L}}$ and frequencies
$\nu_{\rm{L}}$ and $\nu_{\rm{L}}+\delta\nu$ form a moving
light-grating. The common frequency $\nu_{\rm{L}}$ is chosen to be
in the optics domain but far detuned from atomic resonances to
avoid spontaneous emission. A two-photon transition, involving
absorption of a photon from one beam and stimulated re-emission
into the other beam, results in a coherent transfer of momentum
${\bf p}_{\rm{f}}-{\bf p}_{\rm{i}} =2\hbar{\bf k}_{\rm{L}}$ from
the light field to the atoms, where ${\bf p}_{\rm{i}}$ and
${\bf p}_{\rm{f}}$ are the initial and final momenta of the atoms.
Conservation of energy and momentum leads to the resonance
conditions $E_{\rm{f}} = E_{\rm{i}}+ h \delta\nu$, where (in free
space) the initial and final energies of the atoms are given by
$E_{\rm{i}}=p_{\rm{i}}^2/2m$ and $E_{\rm{f}}=p_{\rm{f}}^2/2m$
respectively. Bragg scattering can be used for different types of
matter-wave manipulation, depending on the pulse length $\tau$.
Using a short pulse ($\tau < 100\,\mu$s), the Bragg beams are
sufficiently frequency broadened that the Bragg process is
insensitive to the momentum distribution within the condensate:
the resonance condition is then satisfied simultaneously for the
entire condensate. If the Bragg laser power and pulse duration are
then selected to correspond to the $\pi/2$ condition, the
probability of momentum transfer to the atoms is $50$ percent:
this is a 50/50 beam splitter for the condensate, between two
different momentum states. When using longer pulses (for example
$\tau = 2\,$ms in \cite{Richard2003}), the Bragg process is
velocity selective, and one can apply this technique to momentum
spectroscopy \cite{stenger:1999,Richard2003}.

\begin{figure}
\begin{center}
\includegraphics[width=3.5in]{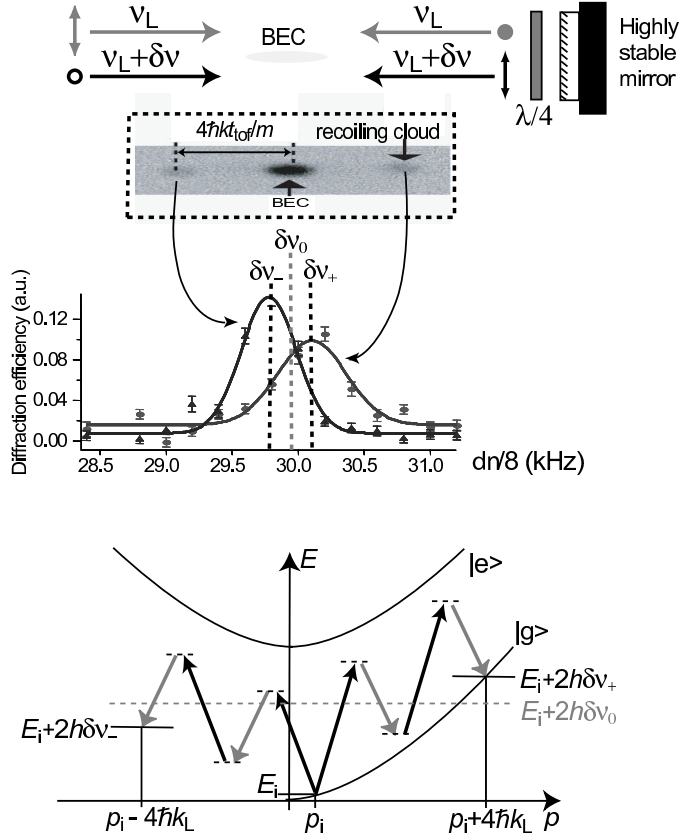}
\end{center}
\caption{Principle of our four photon, dual direction Bragg
scattering scheme. Top: schematic of the experimental apparatus.
Two retro-reflected laser beams form two standing waves of
orthogonal polarisations, moving in opposite directions. Middle:
normalized number of atoms diffracted into each of the two output
channels as a function of Bragg detuning $\delta \nu$. (Inset:
typical absorption image after Bragg diffraction and free
evolution during a time $t_{\rm tof}$.) Bottom: schematic of the
4-photon Bragg resonance condition. For zero initial momentum, the
resonance condition is fulfilled by both standing waves for a
detuning $\delta \nu_0$. For non-zero initial momentum
$p_{\rm{i}}$, the resonance frequency is equally and oppositely
shifted for each of the two channels.
 }\label{fig:Bragg4photon}
\end{figure}

%Expt details
By carefully re-analyzing the Bragg-spectroscopy data of
\cite{Richard2003}, we have extracted a measurement of the ratio
$h/m$ of the Planck constant $h$ to the atomic mass $m$ of
$^{87}$Rb. The experimental sequence proceeds as follows: a
laser-cooled sample of $^{87}$Rb atoms is magnetically trapped in
the 5S$^{1/2}$ $|F=1,m_F=-1\rangle$ state \cite{interrupted} and
then evaporatively cooled to quantum degeneracy. The magnetic
trapping fields are switched off and the atoms fall for 25\,ms.
During this free-fall period, the coherent Bragg-scattering
``velocimeter'' pulse is applied. In this experiment, the
implementation of Bragg scattering is as follows:
%in  \cite{Richard2003} allows directly for extracted the recoil energy
%$\hbar k_{\rm L}^2/2m$:
two orthogonally polarised, co-propagating laser beams of
frequencies $\nu_L$ and $\nu_L+\delta\nu$ and wave vector
${\bf k}_{\rm{L}}$ are retro-reflected by a highly stable mirror,
with $90^{\circ}$ polarisation rotation (see Figure \ref{fig:Bragg4photon}). With this
scheme, the atoms are subject to two  standing waves moving in
opposite directions and with orthogonal polarisations. In
addition, the relative detuning $\delta\nu$ is chosen so as to
fulfill the second-order (4-photon) resonance condition. This four
laser Bragg-scattering scheme produces a coherent transfer of
momentum of $+4\hbar{\bf k}_{\rm{L}}$ and
$-4\hbar{\bf k}_{\rm{L}}$. This scheme enables us to reject the
effect of a non-zero initial velocity, which can arise from
imperfections in the magnetic trap switch-off. For an initial
velocity $p_{\rm i}/m$, the 4-photon resonance conditions for the
two oppositely moving standing-waves are $\delta\nu_{+} =
\delta\nu_0(1 + p_{\rm {i}}/2 \hbar k_{\rm{L}})$ and
$\delta\nu_{-} = \delta\nu_0(1 - p_{\rm i}/2 \hbar k_{\rm{L}})$
where $\delta\nu_0$ is the Doppler-free value, $\delta\nu_{0} =
(8/2\pi) (\hbar k_{\rm{L}}^2 /2 m)$ (see Figure
\ref{fig:Bragg4photon}). Scanning the Bragg scattering efficiency
in the two directions as a function of $\delta \nu$ yields two
peaks with widths corresponding to the condensate momentum width,
centred at each of the resonance frequencies, $\delta\nu_{+}$ and
$\delta\nu_{-}$ (Figure \ref{fig:Bragg4photon}). After fitting
each individual spectrum with a gaussian distribution, we can
extract the two center frequencies  $\delta\nu_{\pm}$. To correct
the data for the non-zero initial velocity, we then re-center both
spectra around the average value $\delta\nu_0=( \delta\nu_{+} +
\delta\nu_{-})/2$.

%

% [ Rejection of initial velocity ]
%

\begin{figure}
\begin{center}
\includegraphics[width=3.5in]{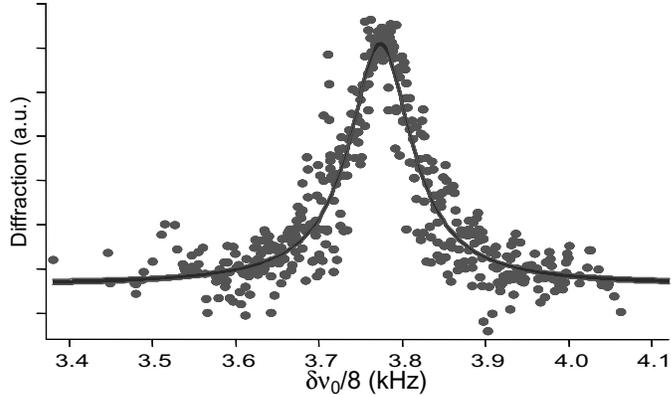}
\end{center}
\caption{Final spectrum (corrected for Doppler effect). The fit to
this spectrum yields the centre frequency $\delta\nu_0$, from
which we obtain the ratio $h/m$.}\label{fig:Corrected_Spectrum}
\end{figure}

% [ Results of h/m measure, interpretation ]780.2462916\times10^{-9}
%expt accuracy of laser lock to 1MHz gives error of 2 10^-15.
%
Using this method and averaging over 350 spectra (Figure \ref{fig:Corrected_Spectrum}), we
find $\delta \nu_0 = 30.189 (4) $ kHz where the figure in
parentheses is the 68\% confidence interval of the fit. We then
deduce a value $h/m \equiv \lambda^2\times\delta\nu_0/4 = 4.5946
(7)\times 10^{-9} \,\rm{J.s.kg}^{-1}$ where the wavelength
$\lambda= 780.246291(2)\times10^{-9}$ of the Bragg beams, slightly
detuned from the $\left(5^2 \rm{S}_{1/2},F=2\right)\rightarrow
\left(5^2 \rm{P}_{3/ 2},F=3\right)$ optical transition, is very
accurately known from \cite{Ye96,Bize99}. The offset between our
measurement and the CODATA value of $h/m$ ($4.59136 \times 10^{-9}
\,\rm{J.s.kg}^{-1}$) can be explained by two major systematic
effects. First, as described in \cite{Richard2003}, the
frequencies $\nu_L$ and $\nu_L+\delta\nu$ of the Bragg scattering
beams were obtained by using two independently driven
acousto-optical modulators (AOM) of center frequency 80 MHz. The
frequency difference $\delta\nu$ was then deduced from the
measurement of the frequency of each AOM driver with a high
precision frequency meter. The reference oscillator used  in
the frequency meter was later characterized to have an accuracy
of about $4\times 10^{-7}$, giving a $\pm 16$ Hz inaccuracy in
the actual frequency difference $\delta \nu$. The resulting
systematic error in our measurement then gives $h/m = 4.5946 (20)
(7)\times 10^{-9} \,\rm{J.s.kg}^{-1}$.  The second systematic
effect is a collisional shift due to interactions in the high
density atomic cloud.  In the following, we will show that this
accounts for the remaining offset.

% [ effects of interactions]
%
Indeed, ultra-cold $^{87}$Rb atoms have
repulsive interactions which modify the Bragg-scattering resonance
condition. The energy of an atom in the condensate is $E_{\rm{i}}
= p_{\rm{i}}^2/2m + Un({\bf r})$. The second term is the
condensate interaction energy: $n({\bf r})$ is the local atomic
density of the condensate and $U=5.147(5) \times
10^{-51}\,$J.m$^3$ is the interaction parameter. Immediately after
Bragg scattering into a different momentum state, an atom
experiences an effective potential $2Un({\bf r})$ due to the
surrounding condensate, and its energy is then
$E_{\rm{f}}=p_{\rm{f}}^2/2m + 2Un({\bf r})$ \cite{stenger:1999}.
We therefore replace the Bragg resonance condition (for zero
initial momentum) with a \emph{local} resonance condition which
takes into account the effect of interactions:
\begin{equation}
    2h\delta\nu_0({\bf r}) = 16 \frac{\hbar^2k_{\rm L}^2}{2m} +Un({\bf r})
    \label{resonance}
\end{equation}
The parabolic density distribution of our Bose-Einstein
condensate, at the moment when the Bragg diffraction occurs, is
\begin{displaymath}
n(x,y,z)=n_0 \cdot \max \left[0 \,;\,
1-(x^2+y^2)/R_{\perp}^2-z^2/{R_z^2}\right]
\end{displaymath}
 with peak density $n_0
\simeq 3.6(4) \times 10^{18} \,$m$^{-3}$ and half-lengths
$R_{\perp} \simeq 9.8\,\mu$m and $R_z \simeq 126\,\mu$m, where $z$
is the direction of the Bragg-scattering. 
Since our measurement of the diffraction efficiency averages over the whole cloud,
the resulting spectrum is then shifted by $U \langle n \rangle/2h \sim 4 U n_{0}/7 $
and broadened. Taking this interaction shift into account, we
correct our measured value of $h/m$:
\begin{eqnarray}
    h/m & = & \lambda^2/4\cdot\left[\langle\delta\nu_0\rangle- U \langle
              n \rangle/2h \right] \nonumber\\
        & \simeq & 4.5939 (21) (7) \times 10^{-9} \rm{J.s.kg}^{-1}.
    \label{h_over_m_final_formula}
\end{eqnarray}
which is in agreement with the CODATA value. Here, the  first
figure in parentheses is the systematic errors discussed
above when we take into account both the frequency calibration inaccuracy and the error on evaluating the
atomic density in the collisional shift. The second figure is the 68\% confidence interval of the fit
determination.

%

% [ interactions are bad... ]
%
The fact that ultra-cold bosons interact is a major drawback for
precision measurements using atom interferometry. As we have seen,
interactions result in a systematic shift as well as a decrease in
measurement precision. In principle, the systematic shifts can be
calculated. However, the interaction parameter $U$ is hard to
measure and is generally not known to better than $\sim 10^{-4}$.
The atomic density is also subject to time fluctuations and is
difficult to know to better than  $\sim 10^{-3}$, reducing the
absolute accuracy.  We have furthermore demonstrated, in an
earlier experiment \cite{Richard2003,Ertmer}, that interactions
produce a loss of coherence of the atomic samples at ultra-low,
finite temperatures, limiting the maximum interrogation time of a
coherent matter-wave atom interferometer.  Finally, even at zero
temperature, the mean-field energy due to interactions is
converted into kinetic energy during free fall, giving rise to a
faster ballistic expansion. This last effect will ultimately
reduce interrogation times.

% [ advantage of interaction-free atomic-sources]
%
From these observations, we conclude that one should ideally use
an interaction-free, ultra-cold atomic source for
ultimate-precision atom interferometry in space. Using bosons, one
could think of two ways of decreasing interaction effects. Close
to a Feshbach resonance \cite{Feshbach}, one can control  the
interaction parameter $U$, which can be made equal to zero for a
certain magnetic field \cite{Feshbach2,Feshbach3}. However
magnetic fields introduce further systematic shifts that are not
controllable to within a reasonable accuracy. Alternatively, one
could try to decrease the density of the sample of atoms, but the
production of large atom number, ultra-low density Bose-Einstein
condensate is a technical challenge not yet overcome
\cite{KetterleTemp2}.

A promising alternative solution is to use quantum-degenerate
fermionic atomic sources \cite{Inguscio}. The Pauli exclusion
principle forbids symmetric 2-body collision wavefunctions, so at
zero temperature a sample of neutral atomic fermions has no
interactions. An ultra-cold fermionic source may still allow very long
interrogation times, even if limited by the excess energy of the Fermi pressure, and 
would therefore be an
ideal candidate for atom interferometry in space with ultimate
precision and accuracy. On-ground experiments using ultra-cold
fermions (Potassium 40) are now under development in our
laboratory and around the world.

% [ inconvenient of fermion sources ]
%
%Unlike bosons, quantum-degenerate ensemble of fermions don't
%exhibit particularly large spatial coherence : even at zero
%temperature, the coherence volume (\emph{i.e.} the coherence
%length to the power three) is equal to the inverse of the atomic
%density, which is much smaller than the atomic cloud size. The
%interferometric principal should therefore not rely on spatial
%coherence.

% [ possible applications]
%
To conclude, we have shown that coherent atomic sources are very
promising for high-precision atom interferometry measurements.
Nevertheless, interactions in quantum-degenerate bosonic gases
cause phase shifts which are difficult to control and will
ultimately limit the measurement accuracy. These shifts could be
overcome either by precise control of the interaction properties,
or by using fermionic, non-interacting samples of ultra-cold
atoms. The use of atom interferometry with ultra-cold sources of
atoms in micro-gravity are of great interest for tests of
fundamental physics in space. There are potentially many
experiments which would benefit from this technology, based on
atomic clocks, interferometers and gravito-inertial sensors. In
space, applications of these include ultra-precise definition of
time, verification of the equivalence principle, measurement of
the fine structure constant $\alpha$ and its drift in time, and
tests of general relativity and post-newtonian gravitation
theories. As an example, the Hyper project \cite{Hyper}, which
aims to measure the Lense-Thirring effect with orbiting
atom-optical gyroscopes, would greatly benefit from using such
coherent sources of cold atoms. In addition, sending an
interplanetary probe equipped with ultra-cold atom interferometer
into space would enable precise mapping of the Pioneer anomaly
\cite{Pioneer}.

%%%%%%%%%%%%%%%%%%%%%%%%%%%%%%%%% acknowledgements %%%%%%%%%%%%%%%%%%%%%
%
The authors would like to thank Arnaud Landragin (BNM/SYRTE) for
his help in the frequency calibration. This work is based on a
re-analysis of the data from \cite{Richard2003} and we would like
to thank especially Fabrice Gerbier and Joseph Thywissen for this
contribution. We would like to acknowledge support from the Marie
Curie Fellowships program of the European Union (J.R.). YLC
acknowledges support from the CNES post-doctoral fellowship
program. This work is part of the CNES supported ICE project with
initial support from D\'{e}l\'{e}gation G\'{e}n\'{e}rale pour
l'Armement, the European Union (Cold Quantum Gases network) and
INTAS (Contract No. 211-855).
%
%%%%%%%%%%%%%%%%%%%%%%%%%%%%%%%%%% bibliography %%%%%%%%%%%%%%%%%%%%%%%%
%

\end{document}